# Non-Hermitian topological phases and skin effects in kagome lattices

Li-Wei Wang[1], Zhi-Kang Lin[1, †] and Jian-Hua Jiang[1, 2, †]

[1]*School of Physical Science and Technology, and Collaborative Innovation Center of Suzhou Nano Science and Technology, Soochow University, 1 Shizi Street, Suzhou 215006, China*

[2]*Suzhou Institute for Advanced Research, University of Science and Technology of China, Suzhou 215123, China*

[†]Correspondence should be addressed to: linzhikangfeynman@163.com (ZKL) and jianhuajiang@suda.edu.cn (JHJ)

**Abstract**

Non-Hermitian physics has added new ingredients to topological physics, leading to the rising frontier of non-Hermitian topological phases. In this study, we investigate Chern insulator phases emerging from non-Hermitian kagome models with non-reciprocal and pure imaginary next-nearest neighbor hoppings. In the presence or absence of $C_3$ rotation symmetry, hybrid topological-skin effects are explored through the identification of distinct corner skin modes in different energy regions within two band gaps. By employing the dynamical analysis, the underlying physics is revealed from the non-Hermitian skin effects associated with the chiral edge states, leading to diverse non-Hermitian bulk-boundary responses. The simplicity of these kagome models and their rich emergent topological phenomena suggest that they are appealing candidates for studying non-Hermitian topological phases. We further discuss the possible realizations of these models in non-Hermitian metamaterials.

## I . INTRODUCTION

The recent studies on the extension of topological phases to non-Hermitian [1-4], nonlinear [5-8], non-Euclidean [9,10], and quantum-optical systems [11-18] have yielded intriguing discoveries beyond the conventional topological physics. In non-Hermitian systems, the unconventional properties such as complex energy spectra and non-orthogonal eigenstates---in extreme cases leading to exceptional degeneracies where the energy eigenvalues and eigenstates of different states coalesce--- enrich the energy bands and Bloch states while posing challenges in characterization of the non-Hermitian topology [19-33]. the presence of non-Hermitian skin effects (NHSEs) fundamentally alters the bulk-boundary responses in non-Hermitian systems [1,3,4,34-

49]. These novel mechanisms and challenges render non-Hermitian topology a captivating frontier of research.

The exploration of non-Hermitian topological phases remains a burgeoning research frontier, particularly for 2- and higher-dimensional systems where the emergent phenomena and their underlying physics are believed to be much richer [43,50-54]. Prototype models are still being sought to provide insight into understanding those non-Hermitian topological systems and offer possibilities for experimental realizations and investigations. Here we study the emergent topological phases in non-Hermitian kagome models with the pure imaginary next-nearest neighbor (NNN) hoppings. By introducing nonreciprocity in the NNN hoppings with and without the $C_3$ rotation symmetry, we unveil several non-Hermitian Chern insulator phases with distinct NHSEs. In particular, the NHSEs are second order, originating from the topological chiral edge states, yielding hybrid skin-topological states at the corner boundaries [51-54]. These results suggest that the non-Hermitian kagome model is an excellent candidate for the study of non-Hermitian topological phases with rich phenomena arising from simple model settings. We further discuss potential realizations of the non-Hermitian kagome model and propose possible experimental studies based on metamaterials.

The paper is organized as follows. In Sec. II. A, we introduce the non-Hermitian kagome model with $C_3$ symmetry, incorporating nonreciprocal hoppings to realize the non-Hermitian skin effect. In Sec. II. B, we investigate the hybrid topological-skin effects [43,50,60-63] in the non-Hermitian kagome model, where the corner skin modes are supported. We also present a physical picture from the dynamical perspective to comprehend the origin of corner skin modes. Namely, the locations of corner skin rely on the combined influence of both positive/negative group velocities and amplification/attenuation of topological edge states. Owing to the $C_3$ rotation symmetry, the corner skin modes are localized at three out of six corners within the supercell. In Sec. III, we study a modified non-Hermitian kagome model without $C_3$ symmetry. The symmetry breaking allows for the existence of excess corner skin modes that are localized at only one or two of six corners, which can be elucidated through the dynamical analysis as well. Finally, we conclude in Sec. IV.

## II. HYBRID SKIN-TOPOLOGICAL MODES

### A. Non-Hermitian kagome model

The non-Hermitian kagome model studied in this work is schematically illustrated in Fig. 1(a). The unit cell consisting of three atomic sites (green color) is outlined by the black hexagonal line. In addition to the nearest neighbor (NN) hoppings (grey lines), the pure imaginary NNN hoppings (bidirectional arrows) are also included, which are incorporated to break the time-reversal symmetry to induce Chern insulator phases. Specifically, the nonreciprocity is introduced in NNN hoppings to open the band gap and bring in non-Hermitian physics. The tight-binding model can be explicitly described by the following Hamiltonian in momentum space,

$$H = H_{NN} + H_{NNN} = \begin{pmatrix} 0 & A_{12} & A_{13} \\ A_{12}^* & 0 & A_{23} \\ A_{13}^* & A_{23}^* & 0 \end{pmatrix} + \begin{pmatrix} 0 & B_{12} & B_{13} \\ B_{21} & 0 & B_{23} \\ B_{31} & B_{32} & 0 \end{pmatrix} \quad (1)$$

where

$$A_{12} = 2t \cdot \cos(\vec{k} \cdot \vec{a}_2), \quad A_{13} = 2t \cdot \cos(\vec{k} \cdot \vec{a}_1), \quad A_{23} = 2t \cdot \cos(\vec{k} \cdot \vec{a}_3) \quad (2a)$$

$$B_{12} = 2\alpha_1 \cdot \cos(\vec{k} \cdot \vec{b}_1), \quad B_{23} = 2\alpha_1 \cdot \cos(\vec{k} \cdot \vec{b}_3), \quad B_{31} = 2\alpha_1 \cdot \cos(\vec{k} \cdot \vec{b}_2) \quad (2b)$$

$$B_{21} = 2\alpha_2 \cdot \cos(\vec{k} \cdot \vec{b}_1), \quad B_{32} = 2\alpha_2 \cdot \cos(\vec{k} \cdot \vec{b}_3), \quad B_{13} = 2\alpha_2 \cdot \cos(\vec{k} \cdot \vec{b}_2) \quad (2c)$$

with the NN and NNN hopping vectors specified by

$$\vec{a}_1 = \left(-\frac{1}{2}, 0\right), \vec{a}_2 = \left(-\frac{1}{4}, -\frac{\sqrt{3}}{4}\right), \vec{a}_3 = \left(-\frac{1}{4}, \frac{\sqrt{3}}{4}\right), \quad (3a)$$

$$\vec{b}_1 = \left(\frac{3}{4}, -\frac{\sqrt{3}}{4}\right), \vec{b}_2 = \left(0, -\frac{\sqrt{3}}{2}\right), \vec{b}_3 = \left(-\frac{3}{4}, -\frac{\sqrt{3}}{4}\right). \quad (3b)$$

Throughout this work, we keep the NN hoppings $t$ as $-1$ and set the lattice constant to unity. The non-reciprocal pure imaginary NNN hoppings $\alpha_1$ and $\alpha_2$ are variable, playing a pivotal role in inducing topological edge states and NHSEs. We focus solely on the hybrid skin-topological modes arising from the topological edge states. To assess whether topological edge states may emerge, we first calculate the real part of the projected band structure as presented in Fig. 1(b), where the NNN hoppings are set to $\alpha_1 = 0.8i$ and $\alpha_2 = -0.6i$ to serve as an example. The corresponding ribbon supercell structure is illustrated in Fig. 2(d). The band structure indeed displays a pair of gapless edge states within both gaps I and II, which are distinguished by different colors depending on which edge they localize. Such the bulk-edge correspondence without time-reversal symmetry can be characterized by the topological invariant known as the Chern number, which remains robust in the presence of non-Hermiticity. According to Ref. [64], the efficient calculation of the Chern number in non-Hermitian systems can be achieved from a real-space perspective, instead of using Bloch or non-Bloch wavefunctions, as briefly recalled in Appendix A. By employing this methodology, we compute the real-space Chern number as functions of $\alpha_1$ and $\alpha_2$, leading to the phase diagrams for both two band gaps, as illustrated in Figs. 1c and 1d, respectively. Notably, both two band gaps within specified parameter regions exhibit a nontrivial Chern number (denoted as $C$) equal to $\pm 1$, signifying the presence of a pair of topological chiral edge states within each gap, which is consistent with the aforementioned projected band structure. Moreover, the Chern numbers exhibit opposite signs in different gaps, indicating that the edge states in different gaps with the same group velocity sign are localized on the opposing edges.

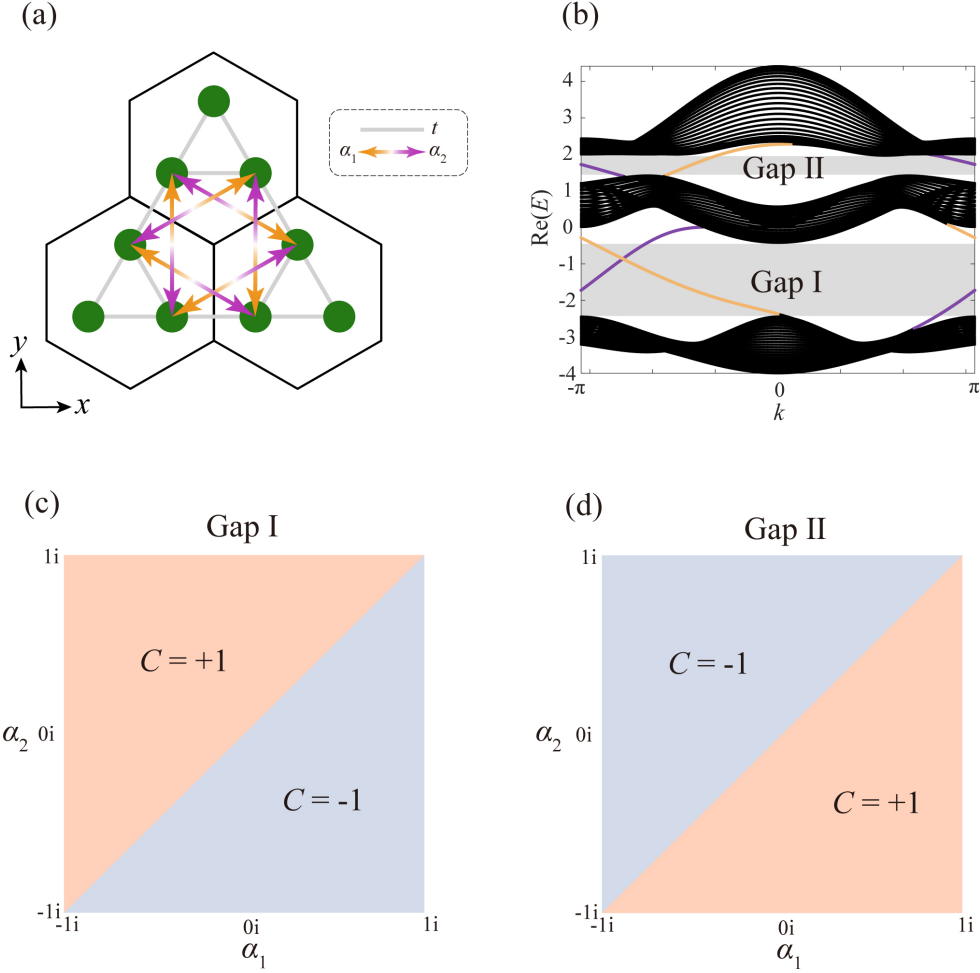

FIG.1. The non-Hermitian kagome model and phase diagrams. (a) The schematic depiction of the non-Hermitian kagome model with $C_3$ rotation symmetry. The black hexagonal represents the unit cell consisting of three atomic sites. The nearest neighbor hoppings are denoted by the grey lines, while the non-reciprocal next nearest neighbor hoppings are indicated by bidirectional arrows in orange and purple colors. (b) The real part of the projected band structure calculated from the ribbon supercell with PBC$x$-OBC$y$. The gaps I and II are marked by grey zones. The edge states localized at the upper (lower) edge are indicated by the purple (orange) curves. The parameters are set as $t = -1$, $\alpha_1 = 0.8i$, and $\alpha_2 = -0.6i$. The number of PBC$x$-OBC$y$ sites is 60. (c) and (d) The real space Chern numbers calculated for gap I and II as functions of $\alpha_1$ and $\alpha_2$. Distinct Chern numbers are represented by different colors.

### B. Chern topological edge states and hybrid skin-topological modes

We now first investigate the non-Hermitian properties of this model in the presence of $C_3$ rotation symmetry. The non-reciprocal NNN hoppings are set as $\alpha_1 = 0.8i$ and $\alpha_2 = -0.1i$, respectively. The complex energy spectra for the non-Hermitian kagome model are displayed in Figs. 2(a)-2(c), respectively, corresponding to three different boundary conditions (PBC$x$-PBC$y$, PBC$x$-OBC$y$, OBC$x$-OBC$y$). PBC (OBC) represents the periodic (open) boundary condition in the $x$- or $y$-direction. The ribbon supercell with PBC$x$-OBC$y$ and the regular hexagonal supercell with OBC$x$-OBC$y$ are schematically illustrated in Figs. 2(d) and (f), respectively. Note that we prefer using the hexagonal supercell rather than the triangular one since the hexagonal supercell can accommodate both two types of edge configurations, namely the upper and lower edges of the ribbon supercell [see Fig. 2(d) for detailed geometries of two types of edges].

Firstly, we observe that the complex spectra comprising of bulk continuum exhibit negligible variations under three distinct boundary conditions [see Figs. 2(a)-2(c)], indicating nearly no first-order skin effects, which originates from the cancellation of nonreciprocity in the bulk due to the only local gauge flux, as elaborated in Ref. [65]. Then, based on the phase diagram, one can deduce the presence of a pair of chiral edge states within each band gap, as clearly evidenced in both the complex energy spectra in Fig. 2(b) and the real part of projected band structures depicted in Figs. 1(b) and 2(e). Alike in Fig. 1(b), we use the purple (orange) color to distinguish the eigenstates localized at the upper (lower) edge, which facilitates the subsequent analysis of the hybrid skin-topological modes.

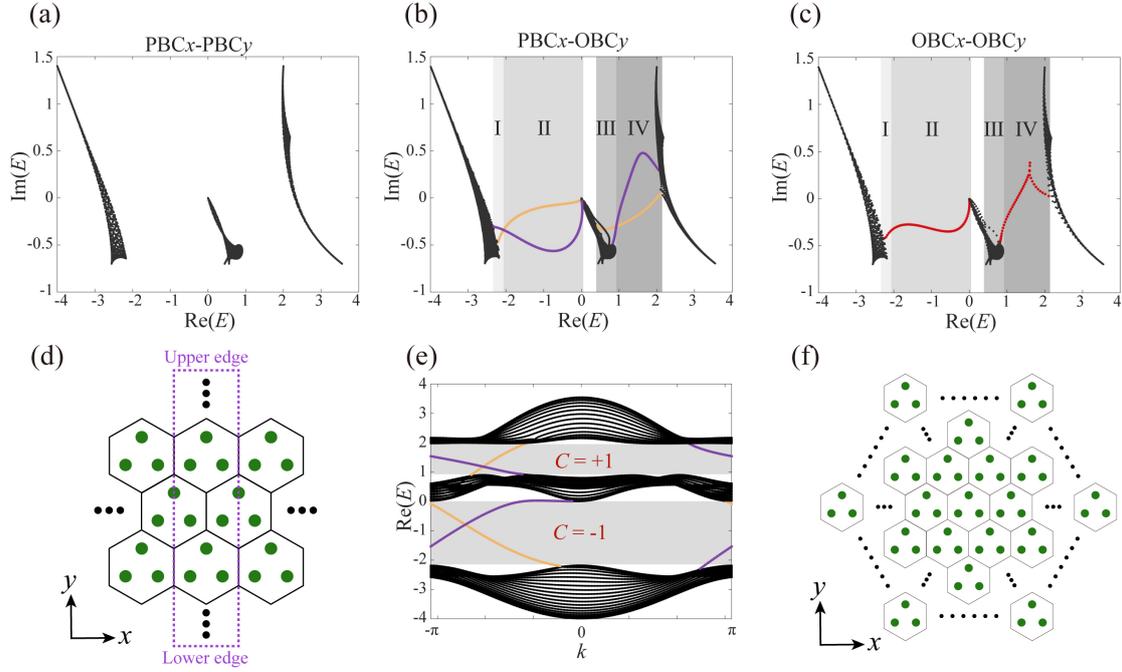

FIG.2. (a)-(c) The complex energy spectra of three distinct boundary conditions, respectively. The corresponding boundary conditions are labeled above each figure. The edge states within both band gaps in (b) are denoted by the purple and orange colors, depending on their localization at the upper and lower edges, respectively. The red dots in (c) represent the hybrid skin-topological corner modes. Each band gap is divided into two regions segmented by the crossing point intersected by two chiral edge states. (d) Schematic depiction of the ribbon supercell with PBC$x$-OBC$y$, as outlined by the purple lines. (e) The real part of the projected band structure alike to Fig. 1(b) but with $\alpha_1 = 0.8i$ and $\alpha_2 = -0.1i$. The Chern number for each band gap is labeled. (f) Schematic depiction of the hexagonal supercell with OBC$x$-OBC$y$. The black dots in both (d) and (f) represent omitted unit cells. The number of PBC$x$-OBC$y$ and OBC$x$-OBC$y$ sites are 60 and 3423, respectively.

Furthermore, it is found that the complex energy spectrum presented in Fig. 2(b) exhibits a crossing point intersected by two edge states within each band gap, which indicates the sign exchange of their imaginary energy difference across two sides of these intersections. Remember that the imaginary part of energy is related to the lifetime of excitation. Prospectively, we divide each band gap into two regions according to the crossing points, as labeled by region I and II for gap I and region III and -IV for gap II shown in Figs. 2(b) and 2(c). More interestingly, as depicted in Fig. 2(c), the complex energy spectrum with OBC$x$-OBC$y$ hosts in-gap states within both two real energy gaps, as highlighted in red color. In particular, these in-gap states reside within the imaginary energy gap (both imaginary and real parts) formed by two topological chiral edge states [range between orange and purple curves in Fig. 2(b)]. We further find that all of these in-gap states are localized at corners, but with distinct wavefunction profiles for different regions, as exemplified in Figs. 3(e)-3(h). We remark that these in-gap corner states are hybrid skin-topological modes that are directly induced by the non-Hermitian skin effect originating from topological chiral edge states.

We now proceed to analyze the origin of the hybrid topological-skin effect. We emphasize that the spectrum of topological chiral edge states within each gap (especially for the gap I) resembles a twisted winding structure hosting two opposite winding numbers, which is famous for bipolar NHSEs in previous studies [66]. However, in our model, we cannot directly utilize the winding number to trace the presence of corner skin modes, as the analogous spectra winding in this work is co-formed by two distinct states localized at different edges. Each edge state does not form a loop due to the chiral anomaly in Chern insulator phases. Although lacking winding numbers, the dynamical analysis introduced in Ref. [3, 4], elaborating the relation between the winding number and the non-Hermitian skin effect, can be directly applied to this work to interpret the hybrid topological-skin effects.

We focus our attention on the crucial characteristic of topological chiral edge states. In addition to considering the imaginary parts of energy derived from the complex spectra depicted in Fig. 2(b), we further incorporate the group velocities of chiral edge states into our analysis, which can be inferred from the real part of the projected band structure presented in Fig. 2(e). Such two factors codetermine the dynamic behavior of chiral edge states. Specifically, the sign of the group velocity is associated with the direction in which a wave packet moves when it resides at that edge. Moreover, the imbalanced lifetime induced by the distinct imaginary-component energy determines either the relative amplification or attenuation of edge states during their dynamic evolution (see Appendix B).

We then take region I as an example to conduct the concrete analysis. It is seen that with the same real part, the energy of a topological edge state marked as orange color exhibits a higher imaginary component compared to the one labeled as purple color [see Fig. 2(b)]. Meanwhile, the group velocities of edge states denoted by purple (orange) color are positive (negative) [see slopes in Fig. 2(e)], indicating the clockwise (anti-clockwise) direction the wave packet propagates in (we define the clockwise direction as the positive direction of the upper edge). Once we apply the OBC in both directions to have a hexagonal supercell, the two types of edges intersect with each other and form six corners. As illustrated in Fig. 3(a), the uppermost, lower-left, and lower-right edges of the supercell correspond to the upper edges of the ribbon supercell, while the other three denote the lower edges.

Based on the above analysis, we can directly derive the directional amplification or attenuation of wave packets at these six edges, as depicted in Figs. 3(a)-3(d) for four different gap regions, respectively. In these figures, the arrows denote the direction of wave packets, while the red (blue) color represents the relative amplification (attenuation) in their respective directions. For instance, in Fig. 3(a), it can be inferred that the right-moving wave packets at the uppermost edge with a longer lifetime will eventually collapse to the upper-right corner after a certain period. Meanwhile, the down-moving wave packets at the upper-right edge with a shorter lifetime will decay exponentially faster. Such dynamics give rise to two types of biases toward different corners and determine the emergent corner skin modes localized exclusively at three out of six corners, as indicated by the luminescent green spheres in Fig. 3(a). Similar analyses are conducted for regions II-IV and their corresponding locations of corner skin modes are displayed in Figs. 3(b)-3(d). To confirm these distinct configurations, we present the wavefunction profile of a representative corner skin mode for each region depicted in Figs. 3(e)-3(h), where the radii of solid circles are proportion to the amplitude strength at each atomic site. It is evident that distributions of these hybrid skin-topological modes align with the prophetic configurations portrayed in Figs. 3(a)-3(d). We emphasize that the locations of corner skin modes for region III (IV) within gap II coincide with region I (II) within gap I due to the reverse signs of both the group velocities and the imaginary-component energy difference between the two edge states. Compared to previous studies [62,63,65], the non-Hermitian kagome model demonstrates hybrid skin-topological modes in both two gaps, thereby expanding the realm of non-Hermitian physics into multi-gap systems. We further specify that no corner skin modes emerge at the energy where the two edge states intersect in complex spectra, as the equivalent imaginary parts of energy lead to no bias toward different corners. In Appendix B, we provide the wavefunction profile of one eigenstate around the cross point in the gap I, which effectively showcases its extension along edges and thus validates our aforementioned analysis.

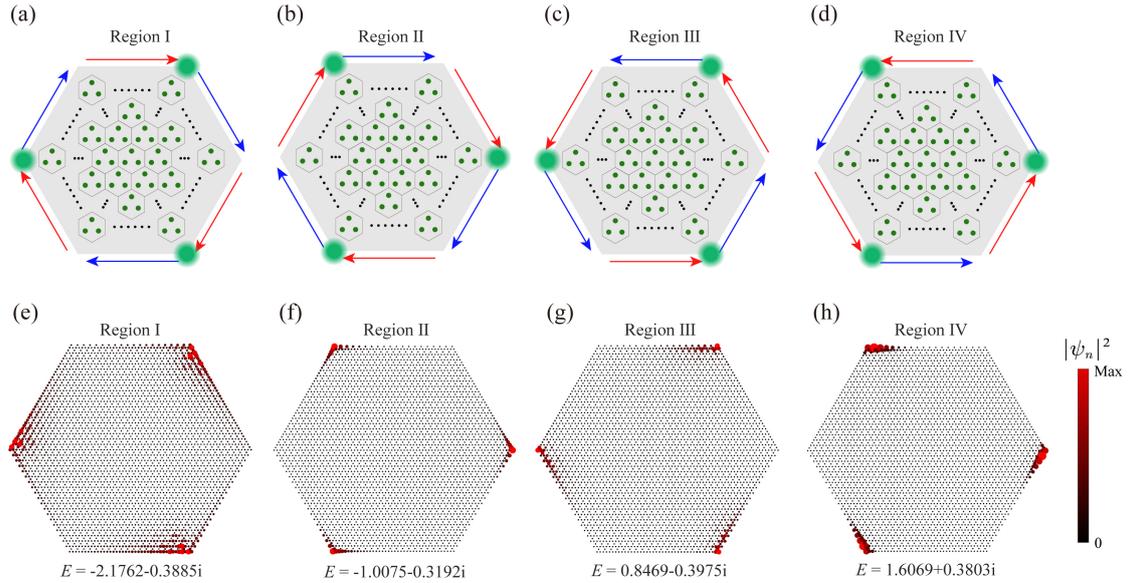

FIG.3. The schematics for corner skin modes in the non-Hermitian kagome model with $C_3$ symmetry. (a)-(d) The schematic diagrams of four configurations for the presence of corner skin modes corresponding to regions I-IV. The red (blue) arrows represent the presence of relative amplification (attenuation) chiral edge states which exhibit wave accumulation in the (opposite) direction to their propagation. Luminescent green spheres represent the potential presence of localized corner skin modes. (e)-(h) The representative hybrid skin-topological modes in the hexagonal supercell with specific eigenvalues corresponding to the configurations in (a)-(d), respectively. The radii of solid circles are proportion to the amplitude strength at each atomic site.

## III. HYBRID SKIN-TOPOLOGICAL MODES WITHOUT $C_3$ SYMMETRY

### A. Modified Non-Hermitian kagome model without $C_3$ symmetry

To further explore corner skin modes, the $C_3$ rotation symmetry is broken by reversing a next-nearest neighbor hopping. As shown in Fig. 4(a), one pure imaginary hopping with a darker color is oriented in the opposing direction, which ultimately results in a modified model possessing neither spatial nor time-reversal symmetry. We proceed in a similar way as in Sec. II. We present in Fig. 4 a representative projected band structure exhibiting edge states and the phase diagrams of Chern numbers. The projected bands are calculated from the ribbon supercell with type 1 PBC$x$-OBC$y$ [there are three different types of mixed boundary conditions due to the absence of $C_3$ rotation symmetry, as depicted in Fig. 5(b)]. The NNN hoppings in Fig. 4(b) are set to $\alpha_1 = 0.8i$ and $\alpha_2 = -0.6i$ as before. We can see that the Chern insulator phases remain robust in the absence of $C_3$ rotation symmetry with the same hopping parameters as in Sec. II.

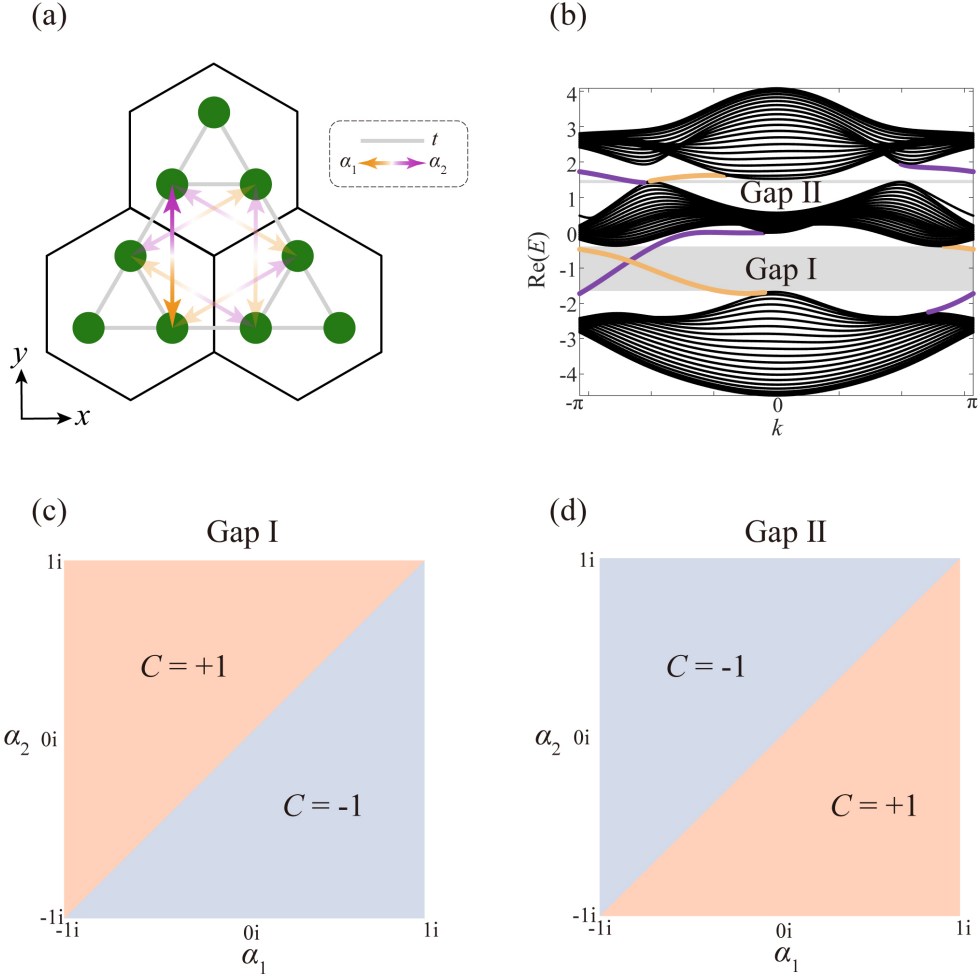

FIG.4. The non-Hermitian kagome model and phase diagrams alike in Fig. 1 but with broken $C_3$ symmetry. (a) The schematic depiction of the non-Hermitian kagome model. A reversed next-nearest neighbor hopping breaking the $C_3$ rotation symmetry is indicated by the darker color. (b) The real part of the projected band structure calculated from the ribbon supercell with PBC$x$-OBC$y$. The parameters are set as $t = -1$, $\alpha_1 = 0.8i$, and $\alpha_2 = -0.6i$. The number of PBC$x$-OBC$y$ sites is 60. (c) and (d) The real space Chern numbers calculated for gap I and II as functions of $\alpha_1$ and $\alpha_2$. Distinct Chern numbers are represented by different colors.

### B. Chern topological edge states and hybrid skin-topological modes without $C_3$ symmetry

We then investigate the complex energy spectra for the modified model with PBC$x$-PBC$y$, OBC$x$-OBC$y$, type 1, 2, and 3 PBC$x$-OBC$y$, as displayed in Figs. 5(a) and 5(c)-5(f), respectively. Therein, the nonreciprocal hoppings are set as $\alpha_1 = 0.8i$ and $\alpha_2 = -0.1i$. Firstly, different from the complex spectra in Fig. 2, we observe that the absence of $C_3$ symmetry brings in discrepancy among the bulk complex spectra with different boundary conditions, which indicates the first-order non-Hermitian effects originating from the nonlocal gauge flux in the bulk [65]. We will not delve into the study of first-order skin modes and focus our attention on the hybrid-topological-skin effects. The complex energy spectra [see Figs. 5(d)-5(f)] and the real parts of projected band structures [see Figs. 5(g)-5(i)] for three different PBC$x$-OBC$y$ exhibit the presence of a pair of chiral edge states within each band gap, but reveal distinct dispersions.

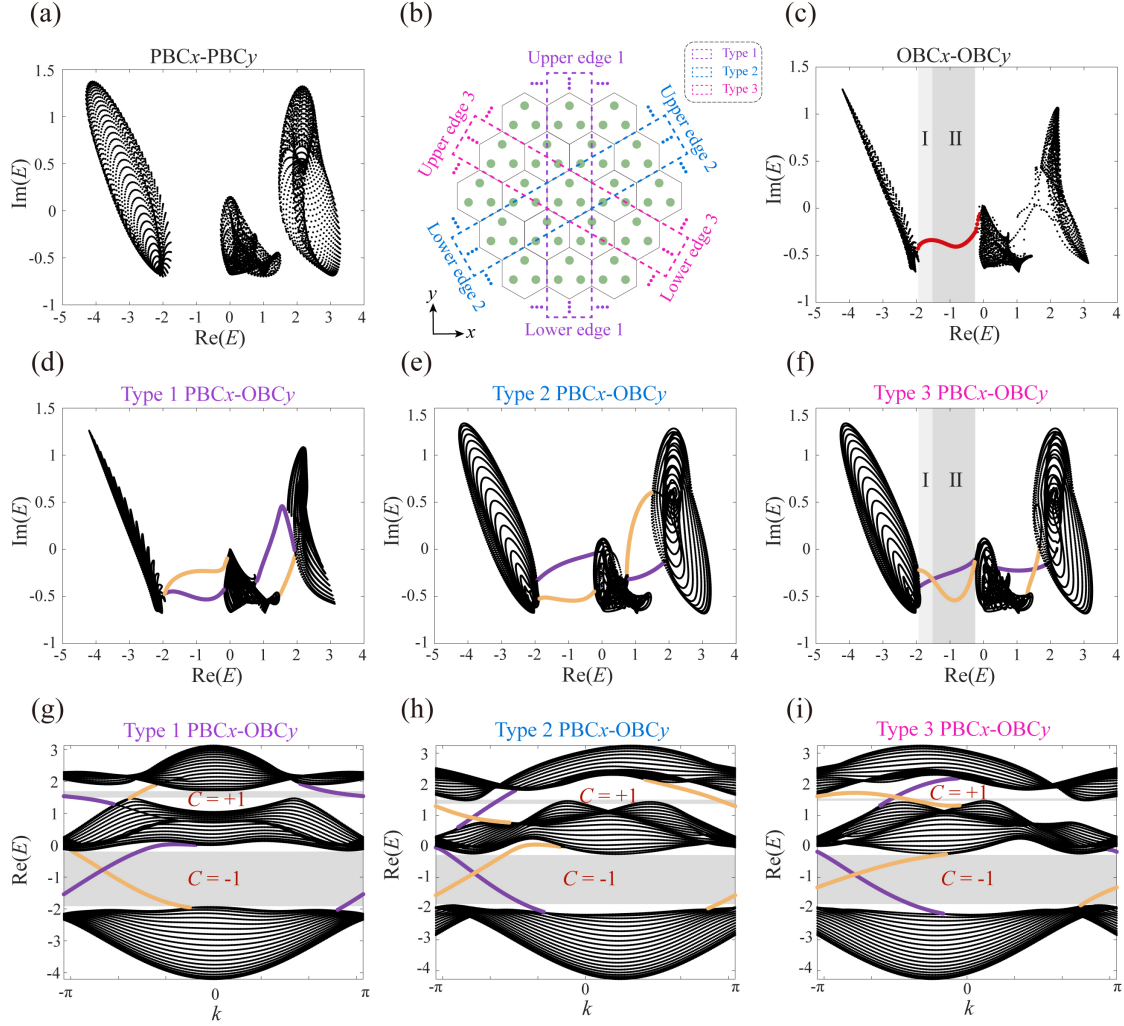

FIG.5. The complex energy spectra and projected band structures with three distinct boundary conditions in the absence of $C_3$ rotation symmetry. (a) and (c) The complex energy spectra with PBC$x$-PBC$y$ and OBC$x$-OBC$y$, respectively. (b) Schematic depiction of the three different types of ribbon supercells with PBC$x$-OBC$y$, as outlined by dashed lines with distinct colors. (d)-(f) The complex energy spectra with three types of PBC$x$-OBC$y$. (g)-(i) The real part of the projected band structures with three types of PBC$x$-OBC$y$. Gaps I and II are marked by two grey areas where the Chern numbers are labeled. The nonreciprocal hoppings are set as $\alpha_1 = 0.8i$ and $\alpha_2 = -0.1i$. The number of PBC$x$-OBC$y$ and OBC$x$-OBC$y$ sites are 60 and 3423, respectively.

Moreover, it is seen that only the complex energy spectrum with type 3 PBC$x$-OBC$y$ exhibits the intersection points of topological chiral edge states within each band gap. We focus solely on the gap I and refrain from further analysis of corner skin modes within gap II due to its tiny band gap. The gap I is divided into two regions according to the crossing points, as labeled by regions I and II shown in Figs. 5(c) and 5(f). Similar to the unmodified non-Hermitian kagome model, the complex energy spectrum with OBC$x$-OBC$y$ hosts corner skin modes within gap I, as highlighted in red color. In particular, these skin modes exist within all the three imaginary energy gaps constituted by two topological chiral edge states [ranges between orange and purple curves in Figs. 5(d)-5(f)]. We further investigate the distinct wavefunction profiles of these corner skin modes, as exemplified in Fig. 6(b) for region I and Fig. 6(d) for region II. Note that four unique wave function distributions of corner skin modes emerge in region II, representing a diverse set of hybrid skin-topological modes resulting from the breaking of the $C_3$ symmetry. We remark that the hybrid skin-topological modes are still directly induced by the non-Hermitian skin effect originating from topological chiral edge states. However, an extension of the previous analysis in Sec. II is necessary to demonstrate the various hybrid skin-topological modes presented here, which are discussed subsequently.

Recall that the dynamic behavior of chiral edge states is jointly determined by the imaginary components of energy and the group velocities associated with them. Once the $C_3$ symmetry is broken, the two factors become protean. The first distinction relies on the fact that the imbalanced lifetimes resulting from distinct imaginary-component energy of two edge states are different for three different types. The second discrepancy stems from the fact that when $C_3$ symmetry is disrupted, the diversity in projected band structures for three types leads to the variation in the magnitude of group velocity for the edge states. Consequently, the traversal time of wave packets localized at three types of edges differs. In conclusion, these two versatile factors exert a substantial influence on the directional amplification or attenuation of edge states during their dynamic evolution, thereby contributing to the diversity of hybrid skin-topological modes (see Appendix B).

We first take region I to perform a similar analysis. In terms of types 1 and 3 (type 2) PBC$x$-OBC$y$, it is apparent that with the same real part, the energy of the topological edge state denoted by purple (orange) color exhibits a higher imaginary component compared to the one represented by orange (purple) color [see Figs. 5(d)-5(f)]. Meanwhile, the group velocities of edge states marked by purple color in type 1 (types 2 and 3) are positive (negative), while those labeled by orange color are opposite [see slopes in Figs. 5(g)-5(i)]. Based on the above analysis, we can graphically demonstrate the directional amplification or attenuation of wave packets at these six edges, as depicted in Fig. 6(a). This configuration leads to a corner skin mode localized only at the lower-left corner indicated by the luminescent green sphere. We then deliberate on the corner skin modes within region II. Regarding types 2 and 3 (type 1) PBC$x$-OBC$y$, the edge states represented by purple (orange) color manifest a higher imaginary component of energy compared to the one labeled by orange (purple) color [see Figs. 5(d)-5(f)]. The signs of group velocities of the edge states are consistent with region I. The corresponding configuration, as illustrated in the left panel of Fig. 6(c), suggests that a corner skin mode is localized at three corners. However, the actual scenario is more intricate including four cases as shown in Fig. 6(d), due to the notable dissimilarities in the imaginary components of energy and the group velocities associated with the edge states at three types of edges. In light of the preceding analysis based on dynamics, a qualitative interpretation is warranted as follows.

We define $P_j \equiv e^{\Delta E_j \cdot t_j}$ to demonstrate the combined effects of the $\Delta E$ and $t$ for every edge indexed by $j$ ($j = 1, \ldots, 6$). The disparity between the imaginary component of energy for each edge state and that of the corner skin mode (defined as the reference energy) is represented as $\Delta E_j$, which serves to characterize the lifetime duration. The time $t_j$ is determined by dividing the unit lattice constant by the group velocity of the edge state with target energy, representing the temporal duration required for a wave packet to propagate a unit lattice distance along the edge. The exponential form of the product $\Delta E_j$ and $t_j$ characterizes the wavefunction profiles at various sites on the edge in dynamics (refer to Appendix B for further elucidation and comprehensive calculation). Assuming a specific corner as the reference starting point and denoting its wavefunction profile as unity, we subsequently apply a sequential product operation to each encountered edge to trace the propagation of the eigenstate wave packet along the six edges of the hexagonal supercell. Finally, by comparing the wavefunction profile magnitudes at each corner, we can assess the manifestation of the hybrid topological skin effect at each corner.

Here, we focus on analyzing the manifestation of the hybrid skin-topological mode of case 1 in region II [first panel in Fig. 6(d)]. Firstly, we calculate $P_j$ marked in Fig. 6(b) for all six edges. Next, we designate the lower-left corner as the reference starting point and assume its wavefunction profile to be unity. Hence, the $P_{ci}$ indicated in Fig. 6(b) at each of three corners labeled by $i$ ($i = 1, 2, 3$) can be calculated. The values of $P_j$, $\Delta E_j$ and $t_j$ for all four cases are listed in Table I in Appendix B. The value of $P_{c1}$ at the starting corner remains at 1, while the $P_{c2}$ and $P_{c3}$ at the other two corners are 0.8428 and 0.8587, respectively. Significantly, during the evolution along the boundary of the hexagonal supercell, the wavefunction profile of the wave packet reaches a maximum at corner 1, thus resulting in the emergence of a localized corner as a corner skin mode [illustrated in the first panel of Fig. 6(d)]. Similar analyses are conducted for cases 2-4 and their corresponding locations of corner skin modes are displayed in the right three panels of Figs. 6(b). We emphasize that as $P_{ci}$ increases, the localization of the corner skin modes becomes more pronounced. Previous studies (Refs. [62, 63]) focus solely on the symmetric models. Here, the introduced concept $P_j \equiv e^{\Delta E_j \cdot t_j}$ presents a more universally applicable approach for analyzing hybrid skin-topological modes, particularly when no symmetry is involved.

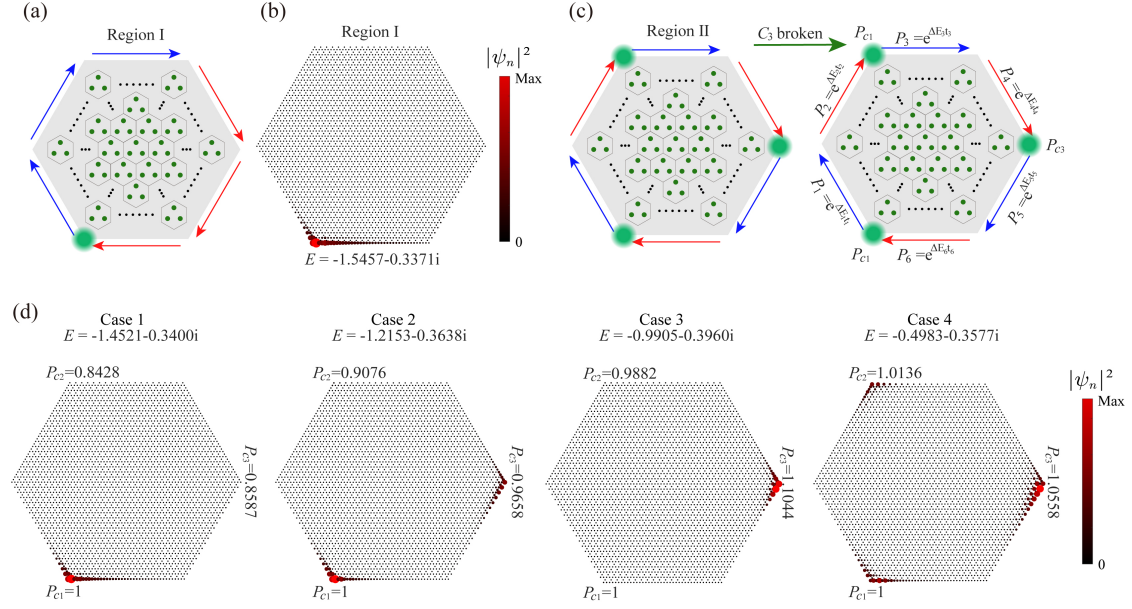

FIG.6. The schematics for corner skin modes in the non-Hermitian kagome model with broken $C_3$ symmetry. (a) and (c) The schematic diagrams of two configurations for elaborating the presence of corner skin modes in regions I and II, respectively. The red (blue) arrow represents the relative amplification (attenuation) of the chiral edge states which exhibit wave accumulation in the (opposite) direction to their propagation. Luminescent green spheres represent the potential presence of localized corner skin modes. (b) and (d) The representative hybrid skin-topological modes with specific eigenvalues for region I and II, respectively. In (c), $t_j$ represents the time required for wave propagation across the entire edge, $\Delta E_j$ signifies the discrepancy in imaginary parts of energy between the edge states and the corner skin modes, while $P_j$ quantifies the magnitude of the combined impact of $\Delta E_j$ and $t_j$, where $j$ ($j = 1, ..., 6$) denotes index of different edges. $P_{ci}$ represents the amplitude of wave function at three potential corners, as illustrated in (c) and (d), where $i$ ($i = 1, 2, 3$) is the corner index. In (b) and (d), the radii of solid circles are in proportion to the amplitude strength at each atomic site.

## IV. Conclusion.

In this research, the time-reversal symmetry of the traditional kagome lattices is initially broken by introducing pure imaginary next-nearest-neighbor couplings, which enables the generation of the topological Chern edge states and the phase transition to non-Hermitian kagome lattices. The presence of hybrid skin-topological modes within both two energy gaps provides empirical validation of our analysis based on dynamics.

Furthermore, we investigated the impact of reversing one of six pure imaginary next-nearest-neighbor couplings, resulting in the breaking of the $C_3$ symmetry in this model. Remarkably, the hybrid skin-topological modes are observed in this modified model as well. Nonetheless, the asymmetry of the edges necessitates the existence of excess corner skin modes that are localized at only one or two of six corners.

The non-Hermitian kagome models with nonreciprocal hoppings considered in this work are exceptionally practicable for experimental investigations. Certainly, the Hermitian counterparts (without nonreciprocal hoppings) have already been experimentally demonstrated in the fields of acoustics and photonics [67-71]. The pure imaginary nonreciprocal hoppings can also be generated by introducing artificial dimensions. In some previous studies [72,73], it has been demonstrated not only for pure imaginary nonreciprocal hoppings but also for complex nonreciprocal ones. This provides a feasible experimental platform for observing hybrid topological skin modes in 2-dimensional non-Hermitian Chern insulators. Furthermore, various 2-dimensional models, such as the square and graphene structures, can be transformed into Chern insulators by introducing complex nonreciprocal hoppings. This enables the exploration of luxuriant hybrid topological skin phenomena, facilitating a more comprehensive investigation. Furthermore, we anticipate that the analytical approach employed in this study, based on the dynamical properties of edge states, can be demonstrated in experimental dynamics. For instance, mechanical systems [74] can provide a formidable tool to observe the dynamic phenomena of non-Hermitian skin modes.

# Appendix

**A. Topological invariants**

The topological invariant serves as a crucial metric for characterizing the inherent topological characteristics of a system without time-reversal symmetry. The non-Bloch Chern number calculated by integrating the Berry curvature over the generalized Brillouin zone is a well-established method to discern the Chern number in non-Hermitian systems. However, the acquisition of the generalized Brillouin zone may pose a formidable challenge in complex structures. Here, one can obtain its topological characterization of non-Hermitian Chern insulators without translational symmetry by calculating the real-space Chern number in a finite lattice, which has been successfully used in some previous non-Hermitian studies [75,76].

The real-space Chern number is defined through the Kitaev formula [64]

$$C_{real\,space} = 12\pi i \sum_{j \in A} \sum_{k \in B} \sum_{l \in C} (P_{jk} P_{kl} P_{lj} - P_{jl} P_{lk} P_{kj}). \quad (S1)$$

A circle region with radius $r$ is chosen and its center is set to coincide with the center of the supercell. The circle is cut into three distinct neighboring regions arranged in a counterclockwise order, marked as "A", "B" and "C" [see Fig. S1(a)]. $j$, $k$ and $l$ mark the sites in A, B, and C regions, respectively. $\hat{P}$ is the projection operator

$$\hat{P} = \sum_{E_n < E_f} |\phi_{n,R}\rangle\langle\phi_{n,L}|, \quad (S2)$$

where $E_f$ is Fermi energy, $n$ is the index of eigenstates. $|\phi_{n,R}\rangle$ and $\langle\phi_{n,L}|$ are the right and left eigenvectors of the supercell.

The real space Chern numbers, denoted as $C(E_f)$, are calculated by varying $E_f$ for the non-Hermitian kagome model with and without $C_3$ symmetry and are presented in Figs. S1(b) and S1(c), respectively. Meanwhile, other parameters are selected as $t = -1$, $\alpha_1 = 0.8i$ and $\alpha_2 = -0.1i$. Notably, two distinct plateaus are observed in both Figs. S1(b) and S1(c), which correspond to the energy ranges associated with two energy gaps. Within these gap regions, the real space Chern number remains quantized as an integer. The real space Chern number is $-1(+1)$ in gap I (II), indicating the presence of a pair of topological edge states in both gap I and II.

It is noteworthy that the real space Chern numbers possess broad applicability across diverse scenarios, particularly in systems lacking translational symmetry, such as the non-Hermitian kagome model under consideration here, quasicrystal [77], and fractal lattice systems [78].

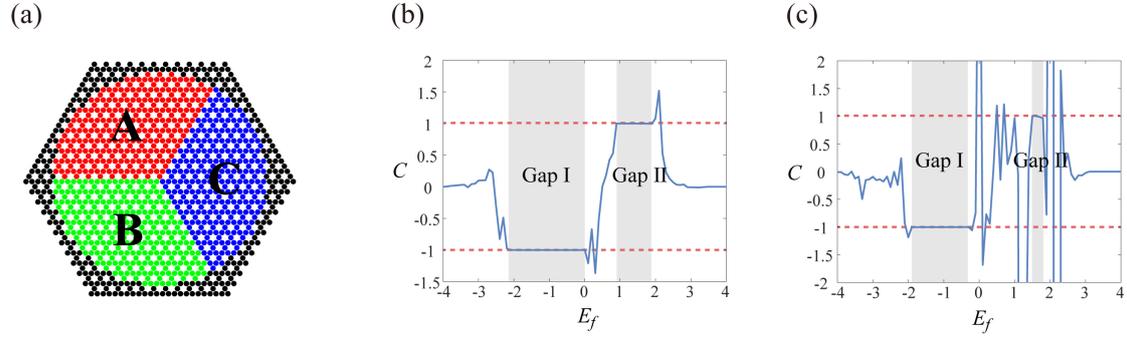

FIG. S1. (a) Schematic diagram of the supercell partitioning when calculating the real-space Chern number. (b) (c) The real-space Chern number as a function of the reference energy $E_f$ (i.e. Fermi energy) for the non-Hermitian kagome model with and without $C_3$ symmetry, respectively.

## B. The phenomena of directional amplification and attenuation of wave packets during eigenstate's dynamic evolution.

In a non-Hermitian system, energy can be represented as a complex number. When the imaginary component of energy is positive (negative), the associated eigenstate will exist an amplification (attenuation), presenting a longer (shorter) lifetime over time. It is crucial to emphasize that the concept of lifetime is inherently relative. For example, in cases where the energy of two eigenstates possess negative imaginary components, a lower (higher) imaginary component corresponds to greater (lesser) energy loss, which exhibits a prolonged lifetime or, in other words, relative amplification (attenuation).

The imaginary component of the corner skin mode is defined as the reference energy, due to its emergence from the collective interaction of all edge states. In addition, the motion of a wave packet situated at the boundary will be determined by the magnitude and sign of the group velocity of the projected band structure associated with a specific energy. The combination of these two concepts enables us to ascertain whether the edge state demonstrates directional amplification or attenuation. Consequently, this analysis enables us to precisely identify the specific corners within the hexagonal supercell structure described in the main text where hybrid skin-topological modes manifest, particularly within non-Hermitian kagome models in the absence of $C_3$ symmetry.

Here, we will present a comprehensive overview of the directional amplification or attenuation during the dynamic evolution of eigenstates. At any given moment, the final state $\Psi(t)$ can be obtained by applying the time evolution operator $U \equiv e^{-iH_{OBC}t}$ to the initial state $\Psi(0)$.

$$\Psi(t) = e^{-iH_{OBC}t} \cdot \Psi(0). \tag{S3}$$

The time evolution operator $U \equiv e^{-iH_{OBC}t}$ can be expanded using the left eigenvectors $\langle \varphi_{j,L}|$ and right eigenvectors $|\varphi_{j,R}\rangle$ as follows:

$$U \equiv e^{-iH_{OBC}t} = \sum_j |\varphi_{j,R}\rangle e^{-iE_j t} \langle \varphi_{j,L}|, \tag{S4}$$

$j$ represents the degrees of freedom of the Hamiltonian $H_{OBC}$. The right eigenvectors $|\varphi_{j,R}\rangle$ and left eigenvectors $\langle \varphi_{j,L}|$ possess biorthogonality.

$$\langle \varphi_{i,L}|\varphi_{j,R}\rangle = \delta_{i,j}. \tag{S5}$$

Therefore, Eq. (S3) can be rewritten as:

$$\Psi(t) = \sum_j |\varphi_{j,R}\rangle e^{-iE_j t} \langle \varphi_{j,L}|\Psi(0)\rangle. \tag{S6}$$

Separating the real part $Re(E_j)$ and the imaginary part $Im(E_j)$ of the complex energy $E_j$, we have:

$$\Psi(t) = \sum_j |\varphi_{j,R}\rangle e^{Im(E_j)\cdot t} e^{-i\cdot Re(E_j)\cdot t} \langle \varphi_{j,L}|\Psi(0)\rangle. \tag{S7}$$

When $E_j$ is a complex number, it leads to a situation where the magnitude of $e^{Im(E_j)\cdot t}$ (defined $P \equiv e^{Im(E_j)\cdot t}$) is not equal to unity. Consequently, the directional amplification or attenuation effects can be determined in the dynamic evolution of eigenstates based on whether $P$ exceeds 1 or not.

Subsequently, we will investigate the emergence of corner skin modes by considering the junction of two edges at an angle, resulting in a corner formed at the point of intersection [Fig. S2(a)]. Let us assume the wavefunction profile of an initial wave packet originates from the position of the $P_L$, possessing a group velocity $v_1$. It propagates through the corner $P_c$ and reaches the right edge, while another wave packet moving with a group velocity $v_2$ to the location of $P_R$. $v_1$ and $v_2$ are determined by the dispersion of the edge states in projected band structures. Considering the equal number of unit cells along both edges (set as unity), we can calculate the time $t_1$ ($t_2$) required for the motion from the left edge (the corner) to the corner (the right edge). The time $t$ is determined by dividing the unit lattice constant by the group velocity of the projected band structure with target energy, representing the temporal duration required for a wave packet to propagate one-unit lattice distance along the edge.

Let $E_+$ ($E_-$) represents the imaginary component of energy at the left (right) edge, the subscript + (-) denotes positive (negative) values according to the reference energy. Therefore, the $P_L$, $P_C$ and $P_R$ can all be calculated and analyzed. By assuming $P_L = 1$, we can calculate that $P_C = e^{E_+ \cdot t1} > 1$, and $P_R = e^{E_+ \cdot t1} \cdot e^{E_- \cdot t2} = P_C \cdot e^{E_- \cdot t2} < P_C$. Consequently, the wavefunction profile at corner C reaches a maximum value $P_C$, indicating the presence of the accumulation of wave packets. Conversely, when $E_+$ ($E_-$) represents the imaginary component of energy at the right (left) edge, we can calculate $P_L = 1$, $P_C = e^{E_- \cdot t1} < 1$, and $P_R = e^{E_- \cdot t1} \cdot e^{E_+ \cdot t2} = P_C \cdot e^{E_+ \cdot t2} > P_C$. The corner does not manifest itself in this particular scenario.

Performing the aforementioned analysis on all six corners within the hexagonal supercell and subsequently calculating the $P_{ci}$ enables us to discern the corners associated with the highest $P_{ci}$. The corner exhibiting the highest $P_{ci}$ value will consistently demonstrate the emergence of skin modes, while other corners will display no or relatively less pronounced profiles, depending on the ratio between their respective $P_{ci}$ and the maximum $P_{ci}$.

In situations where the energy spectra of two edge states intersect, the energy difference in imaginary parts between the two edge states becomes zero, resulting in the fact that $P_j$ is unity along the entire boundary. The wavefunction profiles at this energy exhibit the extension along the boundaries in the hexagonal supercell, as shown in Fig. S2(b) where the energy is chosen at the intersection of two edge states for the first band gap in Fig. 2(b) in the main text.

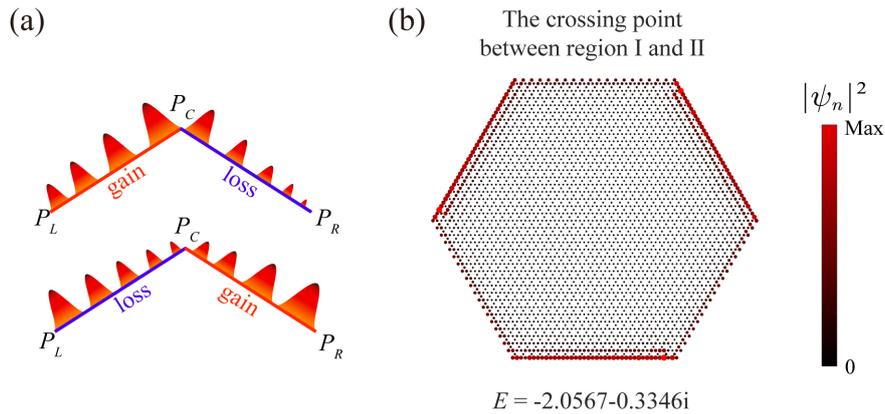

FIG. S2. (a) Schematic diagram of the phenomena of directional amplification and attenuation of wave packets during eigenstate's dynamic evolution. (b) The corner skin mode in the supercell is characterized by the selected eigenvalues, which represent the intersection between region I and II in the non-Hermitian kagome model with $C_3$ symmetry.

Finally, a comprehensive explanation is provided on the determination of parameters $\Delta E_j$ and $t_j$, which is necessary for computing $P_j$ at each edge. Utilizing the reference energy $E_{supercell}$, we can identify complex energies $E_{j,PBCx-OBCy}$ ($j$ is the index of edges) within all types of projected band structures that share the same $Re(E_{supercell})$ value. If $Im(E_{j,PBCx-OBCy}) > Im(E_{supercell})$, the directional amplification of the wave located at this boundary during eigenstate dynamic evolution will be observed, with $Im(E_{supercell})$ serving as the reference energy. We then define $\Delta E_j = Im(E_{j,PBCx-OBCy}) - Im(E_{supercell}) > 0$. Conversely, if $Im(E_{j,PBCx-OBCy}) < Im(E_{supercell})$, the directional attenuation of the wave located at this boundary during eigenstate dynamic evolution will be observed, leading to the definition of $\Delta E_j = Im(E_{j,PBCx-OBCy}) - Im(E_{supercell}) < 0$. This procedure enables us to determine $\Delta E_j$ for each boundary.

Each energy $E_{j,PBCx-OBCy}$ associated with each boundary corresponds to a specific wave vector $k_j$ on the projected band structure. We can determine the group velocity $v_j$, utilizing $v_j = \frac{\partial Re(E(k))}{\partial k}|_{k_j}$. By dividing the number of unit cells at the boundary by the group velocity, we can calculate the time $t_j$ required for the wave packet to propagate across this boundary. It is worth noting that in our two hexagonal supercell models, the number of unit cells at each boundary is uniform, allowing us to simplify calculations by setting it as unity. Henceforth, both $\Delta E_j$ and $t_j$ for each boundary can be obtained. Subsequently, we apply the aforementioned analytical approach to identify the precise locations where the corner skin modes become apparent. For reference, we calculate values of $P_j$, $\Delta E_j$ and $t_j$ for four cases for the non-Hermitian kagome model with broken $C_3$ symmetry, as presented in Table I.

| Case 1 | | Case 2 | |
|---|---|---|---|
| $P_j$ | (0.8107, 1.0395, 0.8802, 1.1576, 0.9843, 1.0977) | $P_j$ | (0.8237, 1.1019, 0.8740, 1.2174, 0.8405, 1.1526) |
| $\Delta E_j$ | (-0.1817, 0.0316, -0.1231, 0.1396, -0.0105, 0.0687) | $\Delta E_j$ | (-0.1714, 0.0869, -0.1282, 0.2066, -0.0967, 0.1203) |
| $t_j$ | (1.1548, 1.2279, 1.0362, 1.0480, 1.5032, 1.3562) | $t_j$ | (1.1316, 1.1159, 1.0498, 0.9522, 1.7969, 1.1805) |
| Case 3 | | Case 4 | |
| $P_j$ | (0.8472, 1.1665, 0.8765, 1.2750, 0.7657, 1.1975) | $P_j$ | (0.8400, 1.2067, 0.8113, 1.2840, 0.8309, 1.1461) |
| $\Delta E_j$ | (-0.1474, 0.1463, -0.1212, 0.2694, -0.1343, 0.1641) | $\Delta E_j$ | (-0.1654, 0.1830, -0.1629, 0.2849, -0.0644, 0.1249) |
| $t_j$ | (1.1256, 1.0531, 1.0872, 0.9019, 1.9872, 1.0982) | $t_j$ | (1.0542, 1.0267, 1.2838, 0.8775, 2.8743, 1.0913) |

Table I. The values of $P_j$, $\Delta E_j$ and $t_j$ for four cases for the non-Hermitian kagome model with broken $C_3$ symmetry.